\documentclass[twocolumn]{AASTeX62}
\usepackage{natbib}
\usepackage[section]{placeins}
\usepackage{graphicx}

\begin{document}

\title{Forward Modeling of the Kepler Stellar Rotation Period Distribution: Interpreting Periods from Mixed and Biased Stellar Populations}
\shorttitle{}

\author{Jennifer L. van Saders}
\affiliation{Institute for Astronomy, University of Hawai'i, Honolulu, HI 96822}
\affiliation{Observatories of the Carnegie Institution for Science, Pasadena, CA  91101}
\affiliation{Department of Astrophysical Sciences, Princeton University, Princeton, NJ 08544}
\email{jlvs@hawaii.edu}

\author{Marc H. Pinsonneault}
\affiliation{Department of Astronomy, The Ohio State University, Columbus, OH 43210}

\author{Mauro Barbieri}
\affiliation{Instituto de astronom'a y ciencias planetarias,Universidad De Atacama, Copayapu 485, Copiap', Chile}

\shortauthors{van Saders}

\newcommand{\modfraction}{\ensuremath{62}}
\newcommand{\fracold}{\ensuremath{72}}
\newcommand{\fracyoung}{\ensuremath{4}}
\newcommand{\rothresh}{\ensuremath{2.08}}
\newcommand{\pinrothresh}{\ensuremath{2.02}}
\newcommand{\rocritrothresh}{\ensuremath{2.00}}
\newcommand{\rothreshisrocrit}{\ensuremath{1.98}}
\newcommand{\ropick}{\ensuremath{2.00}}
\newcommand{\detfrac}{\ensuremath{2.7}}
\newcommand{\detfracsolar}{\ensuremath{64}}
\newcommand{\lowampfrac}{\ensuremath{5}}
\newcommand{\lowampfracfast}{\ensuremath{42}}
\newcommand{\soldet}{\ensuremath{68}}
\newcommand{\safeedge}{\ensuremath{5100}}
\newcommand{\ageedgeatsolar}{\ensuremath{4.2}}
\newcommand{\ageedgeatkraft}{\ensuremath{1.5}}
\newcommand{\hotstarblends}{\ensuremath{31}}
\newcommand{\solblendfrac}{\ensuremath{27}}
\newcommand{\coolblendfrac}{\ensuremath{12}}
\newcommand{\sgbraw}{\ensuremath{26}}
\newcommand{\sgbkepselect}{\ensuremath{12}}
\newcommand{\edgeampsol}{\ensuremath{1600}}
\newcommand{\edgeperiod}{\ensuremath{31}}
\newcommand{\popfracampedge}{\ensuremath{10}}
\newcommand{\edgefraclowamp}{\ensuremath{9}}
\newcommand{\edgefraclowampfast}{\ensuremath{18}}
\newcommand{\incdetfrac}{\ensuremath{87}}
\newcommand{\incdetfracedge}{\ensuremath{59}}
\newcommand{\fracunresolved}{\ensuremath{84}}

\newcommand{\kepselectTeff}{\ensuremath{10} }
\newcommand{\kepselectKp}{\ensuremath{0.5} }
\newcommand{\kepselectLogg}{\ensuremath{1.2} }

\begin{abstract}

Stellar surface rotation carries information about stellar parameters---particularly ages---and thus the large rotational datasets extracted from \textit{Kepler} timeseries represent powerful probes of stellar populations. In this article, we address the challenge of interpreting such datasets with a forward-modeling exercise. We combine theoretical models of stellar rotation, a stellar population model for the galaxy, and prescriptions for observational bias and confusion to predict the rotation distribution in the \textit{Kepler} field under standard ``vanilla'' assumptions. We arrive at two central conclusions: first, that standard braking models fail to reproduce the observed distribution at long periods, and second, that the interpretation of the period distribution is complicated by mixtures of unevolved and evolved stars and observational uncertainties. By assuming that the amplitude and thus detectability of rotational signatures is tied to the Rossby number, we show that the observed period distribution contains an apparent ``Rossby edge'' at $\textrm{Ro}_{thresh} = \rothresh$, above which long-period, high-Rossby number stars are either absent or undetected. This $\textrm{Ro}_{thresh}$ is comparable to the Rossby number at which \citet{vansaders2016} observed the onset of weakened magnetic braking, and suggests either that this modified braking is in operation in the full \textit{Kepler} population, or that stars undergo a transition in spottedness and activity at a very similar Rossby number. We discuss the observations necessary to disentangle these competing scenarios. Regardless of the physical origin of the edge, old stars will not be observed at the rotation periods expected under vanilla braking scenarios, biasing the inferred age distributions: this affects stars older than $\sim 9$ Gyr at $T_{eff} = \safeedge$K, older than roughly $\ageedgeatsolar$ Gyr at solar temperatures, and $\ageedgeatkraft$ Gyr at 6500K. Below $\safeedge$K, rotation periods should be detectable and viable age diagnostics even in the oldest stars in the population. Beyond the presence of a long-period edge, the mixed nature of the stellar population further complicates the interpretation of rotation periods. In particular, for stars hotter than $5500$K, $K_p<16$ mag, with rotation periods less than $70$ days, $\sgbraw\%$ of the stars in the raw TRILEGAL population model are subgiant ``contaminants'', although this fraction is reduced to $\sgbkepselect \%$ if the \textit{Kepler} target selection process was efficient at discriminating evolutionary state. These stars may be present at the same colors and periods as their MS counterparts but have a different period-age relationship than dwarfs; their subgiant nature should be apparent with the addition of Gaia astrometric information.
\end{abstract}

\keywords{stars: fundamental parameters,stars: evolution, stars: magnetic field, stars: rotation, stars: solar-type }

\section{Introduction} \label{sec:introduction}
Rotation is one of the fundamental properties of stars, and yet it is only recently that we have datasets with spot-modulation rotation period measurements that number in the tens of thousands, rather than the hundreds. We are poised to study the rotational characteristics of mixed, complex populations, to uncover hitherto unrecognized behaviors, and to utilize rotation as a tool alongside more traditional stellar observables.

Stars are born with a range of rotation periods. These periods evolve over the lifetime of the star as a consequence of angular momentum loss, transport, and evolution of the stellar moment of inertia. A striking dichotomy exists in the observed rotational distribution of solar-type stars: all but the youngest (single) cool stars are slow rotators, with periods of tens of days, whereas the hot stars ($T_{eff} > 6250$K) are rapid rotators, often with periods of a few days \citep{kraft1967}. This behavior is the consequence of deep convective envelopes in cool stars that help to drive magnetic dynamos, which enable angular momentum loss via magnetized stellar winds \citep{parker1958, weber1967}. The hot stars, with very shallow convective envelopes, do not undergo this magnetic braking on the main sequence and largely retain their rapid rotation and wide range of initial rotation periods. Subgiants of both classes of MS rotators should display slower rotation, due both to physical expansion and the presence of deep convective envelopes and the accompanied magnetic braking across all masses \citep{vansaders2013}.

The magnetic braking in cool dwarfs produces a relationship between rotation period and age. Rotation rate is observed to decline roughly as $t^{-1/2}$ \citep{skumanich1972} in young- to intermediate-age main sequences stars. The strong rotation rate dependence of the magnetic braking \citep[typically $dJ/dt \propto \omega^3$, e.g.][]{kawaler1988} causes the wide range of initial rotation rates to converge to a nearly unique value at ages greater than $\sim 0.5 $ Gyr for solar analogs (\citealt{pinsonneault1989}; see \citealt{gallet2015} for a recent discussion including mass trends). The technique of ``gyrochronology" \citep{barnes2007} capitalizes on this behavior to provide rotation-based ages. 

Gyrochronology relationships are generally calibrated empirically, using samples of stars with independently determined ages \citep[e.g.][]{mamajek2008, meibom2009,barnes2010,garcia2014,angus2015}. For sub-solar stellar mass stars, rotation-based age determinations may be among the most precise and practical, since rotation can be more constraining than isochrone methods, and requires only a modest observational investment in comparison to other high-precision methods, such as asteroseismology \citep{epstein2014}. It seems likely that for at least some subset of masses and evolutionary states, period-age relations will be the tool of choice for inferring precise ages for large samples of dwarf stars.  

The \textit{Kepler} mission has provided unprecedented access to the rotation periods of stars thanks to its long duration ($\sim 4$ years), extremely precise ($\sim 10$s of ppm for bright stars), high-cadence (30 minute) observations. Rotating, spotted stars display modulation in their lightcurves, providing a means to measure stellar surface rotation periods. Several groups have extracted large rotation datasets from the \textit{Kepler} lightcurves to date: $\sim1000$ exoplanet candidate host stars \citep{walkowicz2013,ceillier2016,angus2018}, $\sim 300$ asteroseismic dwarfs \citep{garcia2014}, 12,000 FGK stars \citep{nielsen2013}, 24,000 stars in \citet{reinhold2013}, $\sim34,000$ stars \citep{mcquillan2013, mcquillan2014}, and $\sim 18,500$ stars in \citet{reinhold2015}. We are therefore in a position, for the first time, to study the rotation distributions of large samples of stars drawn from a full and varied stellar population.  

\textit{Kepler} data have furthermore provided a rich calibration set for gyrochronology. Observations of the open cluster NGC6819 confirm the viability of gyrochronology for intermediate-aged (2.5 Gyr) solar mass stars \citep{meibom2015}, but asteroseismic data have provided the first evidence that standard gyrochronology relations fail in older stars \citep{angus2015,vansaders2016}. The combination of new calibrators and large field star samples make this an ideal time to address questions regarding the magnetic braking behavior of low-mass stars.

We expect the interpretation of large collections of rotation periods to require care, particularly when it comes to inferring rotation-based ages. The cool, single dwarf population itself contains a mixture of stellar types, ages, and compositions, all of which affect the rotation periods we expect to observe. Furthermore, rotation periods that are drawn from the field are subject to a number of ``contaminating'' populations: hot stars, evolved stars (mainly subgiants), and synchronized or blended binaries. These sub-populations may have periods that are identical to stars in the single, cool dwarf population, and yet obey a very different mapping between period and age \citep{vansaders2013}. In addition to a complex underlying population, most samples of rotation periods should also be subject to detection bias: more rapidly rotating and active stars are easier to detect, which tends to preferentially favor the detection of relatively young objects.   

This is a well-posed problem for forward modeling with full stellar evolutionary models. As discussed in \citet{vansaders2013} and \citet{vansaders2016}, a semi-empirical treatment of angular momentum evolution, rather than parametric fits between color and rotation period, enables a far more nuanced treatment of rotation. These models include empirical initial conditions and a calibration of angular momentum loss rates against cluster and field stars of known mass, composition and age, coupled to evolutionary models that track structural evolution with time. This approach is particularly important for field populations with a mixture of metallicities, as it naturally accounts for the structural impact of composition. It is also important to conserve angular momentum in the absence of torques, which is not a required feature of empirical gyrochronology relations. We therefore perform a forward modeling exercise and compare the results with the observed stellar rotation period distribution in the \textit{Kepler} field. 
 
\begin{figure}
        \centerline{\includegraphics[angle=0,scale = 1.0]{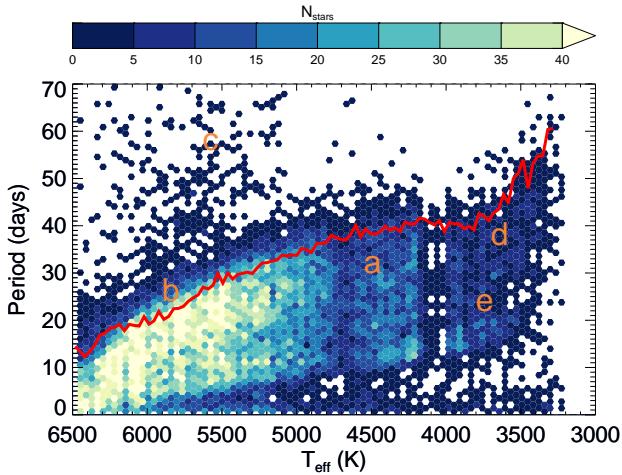}}
        \caption{The observed distribution of rotation periods from MMA14, annotated with several features of note. a: effective temperature catalog artifact,  b: long-period edge, c: long-period overdensity, d: M-dwarf ``dip'', e: bimodal period distribution. The solid red curve represents the $95^{th}$ percentile rotation period, and will be used as a reference throughout this article.}
        \label{MMA14_diagram}
\end{figure}

\section{Rotation and Galactic Population Modeling} \label{sec:vanilla}
A standard model of a rotating stellar population requires a certain minimum subset of assumptions and inputs. Our approach incorporates three distinct components:
\begin{enumerate}
 \item A theoretical model of angular momentum loss in stars. This includes the choice of initial conditions, internal angular momentum transport and loss prescription, and the structural evolution as a function of mass, composition, and age. 
 \item A galactic population model, providing a distribution of the masses, ages, and compositions of stars that should be present in the \textit{Kepler} field.
 \item A model of the \textit{Kepler} targeting and selection function, and selections imposed on the observed \textit{Kepler} stellar sample. 
\end{enumerate}

Once combined, we can construct a model of the period distribution in the \textit{Kepler} field that is directly comparable to the results of searches for periodicity, and assess the performance of our model against the observations.   

\subsection{Observational Comparison Set}

Although multiple teams have developed and utilized pipelines to extract rotation periods from \textit{Kepler} lightcurves, we choose to work with the period sample of \citet[][hereafter MMA14]{mcquillan2014}. This particular rotation dataset represents the largest homogeneous determination of rotation periods in the \textit{Kepler} field to date, and the methodology employed tends to recover longer period variables than competing methods. It has been tested with hound-and-hare injection and recovery exercises \citep{aigrain2015}, and utilizes a set of well-defined criteria, determined via a training set, for distinguishing between periodic and non-periodic sources. We have reproduced the MMA14 $T_{eff}$-period diagram in Figure \ref{MMA14_diagram}, which represents $\sim 34,000$ stars. Features in the rotation distribution of particular note are marked in the diagram. Some features, such as the gap (Fig \ref{MMA14_diagram}, a) at around 4500 K, simply reflect artifacts in the catalog properties (in this case, a mismatch in $T_{eff}$ solutions between cool and hot dwarfs.)  Others, however, represent global properties to be explained by our model. There is a reasonably sharp upper bound (Fig \ref{MMA14_diagram}, b) to the periods for our sample, but for hotter targets there is also a population of relatively long-period stars (Fig \ref{MMA14_diagram}, c). This is important because it indicates that \citet{mcquillan2014} could detect long periods in some targets but did not see them in most. \citet{mcquillan2014} also noted an ``M-dwarf dip'', visible as a kink in the upper envelope of rotation periods at around 3700K (Fig \ref{MMA14_diagram}, d). Furthermore, there is a notable bimodality in the rotation periods of cool stars (Fig \ref{MMA14_diagram}, e), which MMA14 interpreted as structure in the age distribution of nearby stars. 

\subsection{Stellar Models}
We make use of the grid of stellar rotation models created using the Yale Rotating Evolution Code (YREC) from \citet{vansaders2013} and \citet{vansaders2016}. Briefly, these models assume solid body rotation, and draw their initial conditions from the period distributions in young ($< 0.5$ Gyr) open clusters. Non-rotating stellar models ranging in mass from $0.4-2.0 \textrm{M}_{\odot}$ are evolved while tracking stellar parameters, which are then input into a braking law of the form \citep{matt2012,vansaders2013}:

\footnotesize
\begin{equation}
 \displaystyle \frac{dJ}{dt} = \left\{
        \begin{array}{l l}
         \displaystyle f_K K_M \omega \left(\frac{\omega_{crit}}{\omega_{\odot}}\right)^2 \quad \omega_{crit} \leq \omega \frac{\tau_{cz}}{\tau_{cz, \odot}}\\
         \displaystyle f_K K_M \omega \left(\frac{\omega \tau_{cz}}{\omega_{\odot} \tau_{cz, \odot}}\right)^2 \quad \omega_{crit} > \omega \frac{\tau_{cz}}{\tau_{cz, \odot}}\\
        \end{array} \right.
        \label{eqn:wind_law}
\end{equation}
\normalsize
 where $f_K$ is a scale factor tuned to reproduce the solar rotation period at the solar age, $\omega_{crit}$ the saturation threshold, and $\tau_{cz}$ the convective overturn timescale, calculated using convective velocities inferred from mixing length theory one pressure scale height above the convective boundary. $K_M$ is a collection of scalings with fundamental parameters designed to trace magnetic field strength and stellar mass loss:
\footnotesize
\begin{equation}
   \displaystyle \frac{K_M}{K_{M,\odot}} = c(\omega)\left(\frac{R}{R_{\odot}}\right)^{3.1} \left(\frac{M}{M_{\odot}}\right)^{-0.22} \left(\frac{L}{L_{\odot}}\right)^{0.56} \left(\frac{P_{ph}}{P_{ph,\odot}}\right)^{0.44}, 
\end{equation}
\normalsize
with luminosity $L$, mass $M$, radius $R$, and photospheric pressure $P_{ph}$. Such a form assumes that the mean field strength scales with the photospheric pressure, and that mass loss scales with the X-ray luminosity \citep[e.g.][]{wood2002} which in turn scales with bolometric luminosity \citep[e.g.][]{pizzolato2003}. The factor $c(\omega)$ is the centrifugal correction from \citet{matt2012}; here we assume $c(\omega) = 1$, appropriate for slow rotators. This braking law preserves the transition in rotation rates between hot and cool stars particularly well in comparison to more traditional prescriptions \citep[such as][]{kawaler1988}, which tend to predict slow rotation in the hot stars. In this paper we adopt the calibration of \citet{vansaders2013}, where the initial condition was the deuterium burning birthline. Their best fit was an initial rotation period of 8.134 days, critical rotation velocity $\omega_{crit} = 3.394 \times 10^{-5}\textrm{s}^{-1}$, disk locking timescale of 0.281 Myr, and $f_K = 6.575$. Because the angular momentum evolution is derived in a post-processing fashion from the evolutionary tracks, these models do not include the effects of radial or latitudinal differential rotation, or feedback on the stellar structure or lifetimes due to rotation. The models of \citet{vansaders2016} cover a range of compositions, ranging from $\textrm{[Fe/H]} = -0.4$ to $\textrm{[Fe/H] = +0.4}$ in steps of 0.1 dex, and encompass the majority of stellar metallicities we expect to encounter in \textit{Kepler} stars. We use a simple chemical evolution model of the form $\Delta Y / \Delta Z = 1.0$ to set the helium abundance in our models as a function of metallicity. A detailed description of the input physics in our models can be found in \citet{vansaders2013} and \citet{vansaders2016}.

This choice of braking model represents only one of the many available in the literature, although it is standard in its construction: it is calibrated to reproduce the period distributions in young clusters and the Sun. The bulk stellar population in the \textit{Kepler} field is expected to be old. In the TRILEGAL model, subject to the \textit{Kepler} sample selection biases discussed in Section \ref{sec:selection}, \fracold\% of the stars are between 2-10 Gyrs old. In contrast, only $\sim$\fracyoung\% of stars are younger than 0.5 Gyr. Many of the complications in the modeling of angular momentum loss and evolution are in the treatment of young stars: they display a spread in initial conditions, saturation in the magnetic braking at rapid rotation rates \citep[][]{krishnamurthi1997}, and undergo core-envelope decoupling \citep[see][]{denissenkov2010}. Furthermore, the rotation distributions of young rapid rotators can be biased by contaminating populations such as synchronized binaries and blends with high-amplitude classical pulsators. In comparison, old stars have converged onto rotational sequences that are insensitive to these early physical phenomena; all literature braking laws calibrated on open clusters and the Sun tend to predict more or less the same behavior in old systems \citep[with the notable exceptions of][]{reiners2012, vansaders2016}. This fact motivates our use of \citet{vansaders2013} as a ``vanilla'' braking law.

However, in light of recent results from \citet{angus2015} and \citet{vansaders2016}, we do examine an additional alternate braking law, since it makes very different predictions than standard models in the literature. \citet{vansaders2016} utilizes same braking scheme as \citet{vansaders2013}, with one modification: the prescription includes a critical Rossby number ($\textrm{Ro} \equiv P_{rot}/\tau_{cz}$), $\textrm{Ro}_{crit} = 2.16$, above which angular momentum loss ceases. This modification is designed to reproduce the observed anomalously rapid rotation of old field stars with ages determined via asteroseismology, and results in more rapid rotation at late times than canonical braking formulations. We examine the predicted distributions of rotation periods using both the standard braking law, as well as this modified law which incorporates a critical Rossby number.  

 \subsection{Galactic Stellar Population Model}

In order to construct a reasonable theoretical stellar population for a sight-line through the \textit{Kepler} field, we make use of a TRILEGAL galaxy model \citep{girardi2005} tuned for the \textit{Kepler} field. We ran TRILEGAL using the standard values described in \citet{girardi2005} for the initial mass function, star formation rate and age-metallicity relation functions, the geometric description of the Galaxy, and solar location. Binary stars were not simulated. 

The \textit{Kepler} focal plane is comprised of 21, 5 $\textrm{deg}^2$ CCD modules; during the original \textit{Kepler} mission, the field of view was rotated every 93 days for reorienting the solar arrays. For simplicity, we simulate a 5 $\textrm{deg}^2$ field in TRILEGAL centered on each of the module centers. For each center position, we generate 5 realizations of the stellar population using TRILEGAL, in order to have an ample pool of simulated stars to which we can apply a \textit{Kepler}-like selection function (see Section \ref{sec:selection}).  

In order to effectively match the TRILEGAL model to our stellar rotation models, we make use of only the masses, ages, compositions, positions, distances, and extinctions predicted by TRILEGAL. We then interpolate our stellar model grid to determine all other stellar parameters, including $\textrm{T}_{eff}$, $\log{g}$, rotation period, luminosity, and $\tau_{cz}$. With this procedure, we ensure internal consistency between our predicted periods and stellar parameters, since the assumed physics and chemical evolution present in the input stellar models for TRILEGAL are slightly different from those in YREC. We consider only stellar masses and compositions that fall within our grid boundaries ($0.4 \leq \textrm{M/M}_{\odot} \leq 2.0$, $-0.4 \leq \textrm{[M/H]} \leq 0.4$), which represents $\sim$\modfraction\% of all stars $2.0<\log{g}<6.0$ present in the TRILEGAL model. The majority of the stars in the TRILEGAL model excluded because of grid boundaries are of metallicities lower than $-0.4$; it is worth noting that in the \citet{mathur2017} stellar catalog, only $\sim16\%$ of the catalog stars have metallicities $\textrm{[Fe/H]} <-0.4$. 

We calculate stellar colors and magnitudes using the bolometric corrections and extinctions of \citet{girardi2008} \footnote{available at http://stev.oapd.inaf.it/dustyAGB07/bc/kepler/, http://stev.oapd.inaf.it/dustyAGB07/bc/sloan/, http://stev.oapd.inaf.it/dustyAGB07/bc/ubvrijhk}. We ran the TRILEGAL model with an exponential disk treatment of extinction, with $A_V(\infty) = 0.0378$. Given the effective temperature, surface gravity, and luminosity from the YREC stellar models, in addition to the extinction and distance modulus for each star from the TRILEGAL model, we can estimate the magnitude of each object in the \textit{Kepler} ($K_{p}$) band. 

\begin{figure}
        \centerline{\includegraphics[angle=0,scale = 0.5]{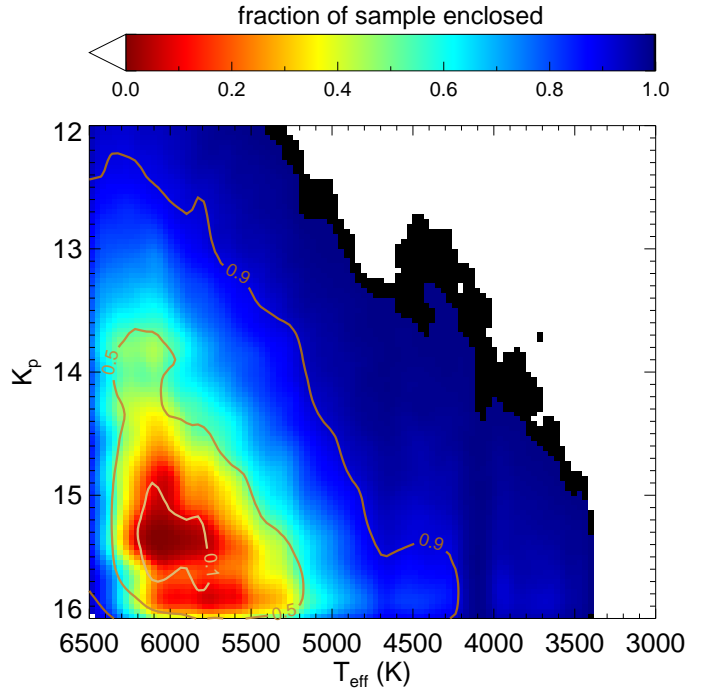}}
        \caption{The density of stars in \textit{Kepler} magnitude and effective temperature. The background represents the density of stars in the TRILEGAL galaxy simulation after we account for the \textit{Kepler} selection bias for FGK stars. Contours denote the fraction of stars (10\%, white; 50\%, tan; 90\%, brown) in the full MMA14 sample enclosed within given regions of magnitude-temperature space.}
        \label{fig:kepselect}
\end{figure}

\subsection{Matching the \textit{Kepler} and MMA14 selection functions} \label{sec:selection}
The \textit{Kepler} mission was designed with the specific intent to search for habitable-zone planets around FGK-type stars, and the target selection and prioritization reflect this fact \citep{batalha2010}. The process of selecting targets for observation accounted for the signal-to-noise of potential transits, number of transits visible during the mission lifetime, minimum observable planet radius, and stellar magnitude. As such, FGK dwarfs are over-represented: hot stars and subgiants are excluded because of their large radii, and M-dwarfs because of their faintness. Rather than reproduce the full target selection exercise here, we instead match the observed distribution of stars in magnitude-$T_{eff}$-log(g) space that were searched for spot modulation in MMA14 \citep[although see][for a discussion of the fidelity of KIC stellar parameter estimates.]{bastien2014}. In practice, we divide all stars observed during the main \textit{Kepler} mission into \kepselectTeff K bins in effective temperature, \kepselectKp mag bins in \textit{Kepler} magnitude, and \kepselectLogg dex bins in surface gravity, using stellar parameters determined in \citet{mathur2017}. We then draw randomly, with replacement, from all stars in all fields and TRILEGAL realizations until we match the number counts in each bin of the observed \textit{Kepler} sample. We use this updated properties catalog as the best approximation to the ``ground truth'' of the distribution of stars that \textit{Kepler} actually targeted. To mimic the selection of MMA14, we also apply the $\textrm{T}_{eff}$ and $\log{g}$ cuts of \citet{ciardi2011} to our model population, and select stars with $K_p < 16$. The selected sample and its agreement with the MMA14 sample is shown in Figure \ref{fig:kepselect}.

\begin{figure}
        \centerline{\includegraphics[scale = 0.45]{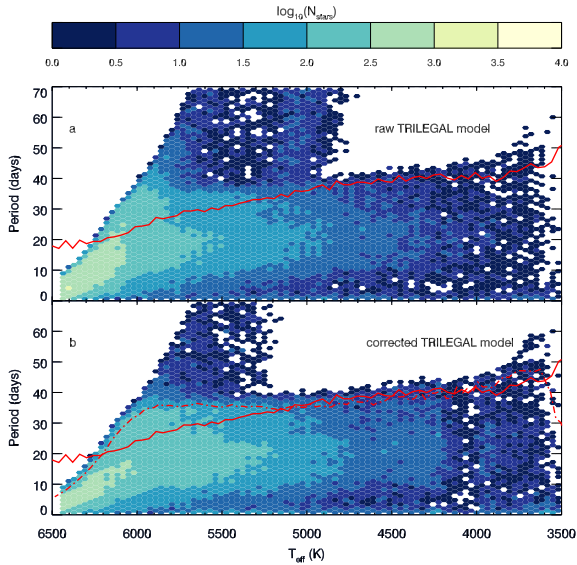}}
        \caption{Distribution of periods in the TRILEGAL model, altered by target selection. Panel a.) shows the distribution of periods as a function of effective temperature predicted using an unaltered TRILEGAL+rotational evolution model, and subjected to a magnitude limit of $K_p<16$. Panel b: The population model subjected to the \textit{Kepler} selection function and MMA14 cuts in log(g) and $T_{eff}$.  In both panels, the solid red line represents the $95^{th}$ percentile period in the observed MMA14 distribution, and the dot-dashed red curve the $95^{th}$ percentile period in the model.}  
\label{page_panel_selection}
\end{figure} 

Figure \ref{page_panel_selection} provides details of how these cuts and selections alter the distribution of rotation periods in comparison to a ``raw'' TRILEGAL model. Panel \textit{a} displays the largely unaltered distribution of stars in period-Teff space from the TRILEGAL subject only to a magnitude cut of $K_p<16$. Panel \textit{b} institutes both a correction for the \textit{Kepler} FGK-star observational bias, and the additional cuts in $T_{eff}$ and logg from MMA14. These selections have the effect of eliminating giants (evident at long periods), a fraction of the hot stars ($>6500$ K), and some subgiants (with core hydrogen $X_c < 0.0002$) from the sample.

\subsection{Lessons from the Vanilla Model}
\begin{figure}
        \centerline{\includegraphics[scale = 0.45]{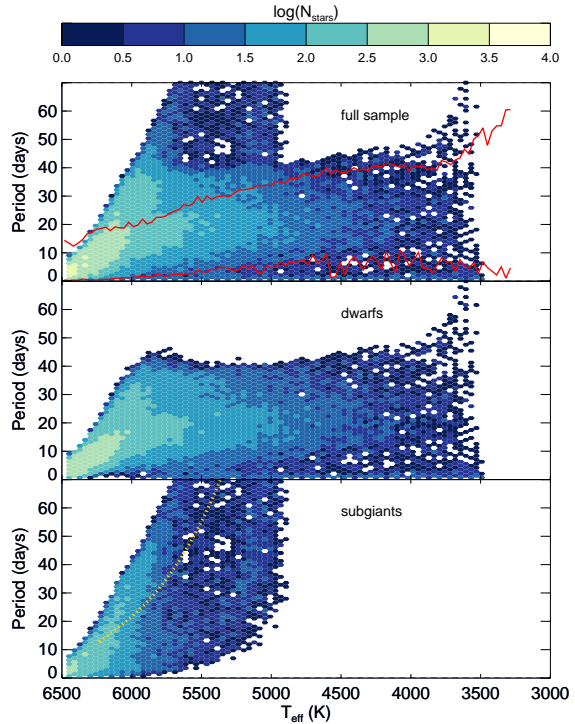}}
        \caption{Distribution of periods in the TRILEGAL model without correction for selection bias. Panel a.) shows the distribution of periods as a function of effective temperature predicted using an unaltered TRILEGAL+rotational evolution model, and subjected to a magnitude limit of $K_p<16$. Panel b: the distribution of periods for dwarfs (defined as having core hydrogen $X_c > 0.0002$). Panel c: the distribution of subgiant stars (defined by having fallen below a core hydrogen fraction of $X_c=0.0002$). The solid red lines represent the $95^{th}$ and $5^{th}$ percentile periods in the observed MMA14 distribution. The yellow dashed line represents the track on the SGB for a star with ZAMS $T_{eff} \sim 6250$K, which roughly demarcates the boundary between those hot stars that did not undergo braking on the main sequence (left of the curve) versus those that did (right of the curve)
.}  
\label{trilegal_intro}
\end{figure}

Panel \textit{a} of Figure \ref{trilegal_intro} displays the rotation distribution obtained by folding together only the rotational models and the TRILEGAL population model for stars with $K_p<16$. Figure \ref{trilegal_intro} provides an overview of the entire population before selection criteria are applied, broken into three groups: cool MS stars (zero-age main sequence (ZAMS) $\textrm{T}_{eff} < 6250$ K), hot MS stars ($\textrm{T}_{eff, ZAMS} > 6250$ K), and subgiants (defined as having core hydrogen $X_c < 0.0002$). This diagram demonstrates that there is a significant diversity, at fixed period, in the rotational histories of stars due to the mixture of masses and evolutionary states. This diversity complicates the interpretation of rotation periods, particularly around solar temperature: a diagnostic of evolutionary state is a necessary ingredient for assigning meaning to a rotation period.

\begin{figure}
        \centerline{\includegraphics[angle=0,scale = 0.45]{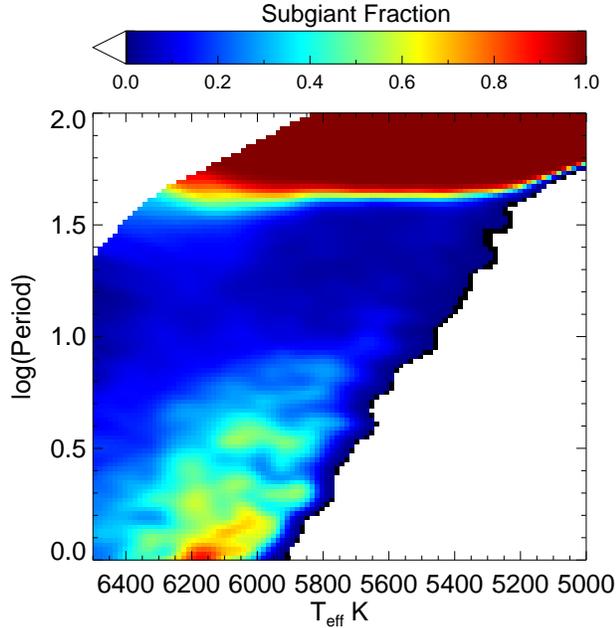}}
        \caption{The fraction of subgiant stars as a function of period and effective temperature. The population model accounts for a magnitude limit, the MMA14 cuts in temperature and surface gravity, and the \textit{Kepler} bias towards observing FGK stars. }
        \label{sgb_fraction} 
\end{figure}

Figure \ref{sgb_fraction} shows the fraction of stars that are on the subgiant branch as a function of effective temperature and period for an FGK-biased (i.e. MMA14-like) sample of stars in TRILEGAL. Two regions of high subgiant fraction are apparent. Stars with temperatures of $5000 \le T_{eff} \le 6000$K at long periods are objects nearing the base of the giant branch. When we examine those 1\% of stars with the longest periods in MMA14, with effective temperatures $5000-5800$K and with flicker-based gravities from \citet{bastien2016} ($\sim 40$ stars), the median surface gravity is $3.74$, indeed indicative of subgiants. The second concentration of subgiants appears at short periods, and represents stars that are born near or above the Kraft break as rapid rotators on the MS, and are now evolving to cooler temperatures along the SGB. Those stars in the MMA14 sample with flicker gravities in this region ($6100<T_{eff}<6300$K, $0.0<\log{P}<0.5$) have a median surface gravity of 3.79, again indicative of subgiants. In comparison, stars in the regions more heavily dominated by dwarfs ($5400<T_{eff}<5800$K, $1.0<\log{P}<1.5$, for example) contain stars with higher gravities (here a median flicker $\log{g}= 4.22$). There are regions of the period temperature diagram that host very pure populations of unevolved stars, such as the cool ($T_{eff}< 5000$K) dwarfs.

In a TRILEGAL population uncorrected for the \textit{Kepler} target selection, subgiants are a significant contaminant: for stars hotter than $5500$K, $K_p<16$ mag $P<70$ days, $\sgbraw\%$ of the stars are evolved. This is higher than the subgiant fraction of $\sgbkepselect \%$ in this domain reported by \citet{mathur2017}; however, in either case the contamination fraction is significant.

\section{Long-Period Behavior}\label{sec:detbias}

The vanilla model was successful at reproducing several features of the observed period distribution: subgiants at long periods and the long-period edge in the cool ($< 5000$K) stars. However, it is evident from Panel \textit{b} of Figure \ref{page_panel_selection} that there are far more long-period stars at solar temperature in the model than are actually observed. We now explore additions to our population model that may account for such a discrepancy.

\subsection{A Detection Edge}

\begin{figure}
        \centerline{\includegraphics[angle=0,scale =1.0]{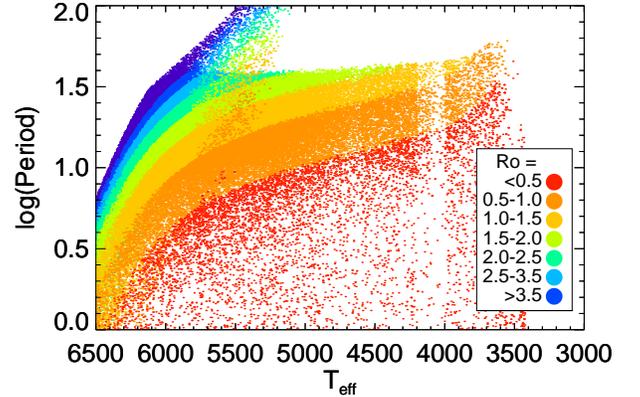}}
        \caption{The model stellar population in period-$T_{eff}$ space, color-coded by Rossby number. The population model assumes a magnitude limit, MMA14 temperature and surface gravity cuts, and accounts for the \textit{Kepler} bias towards FGK stars.}
        \label{ro_patterns}

\end{figure} 

We have thus far made no attempt to address the detectability of rotational spot modulation as a function of spectral type or period in the vanilla model. However, MMA14 found strong trends in the fraction of stars detected in spot-modulation as a function of spectral type, ranging from $\sim0.8$ in the M-dwarfs to only $0.16$ in solar temperature stars. The detectability of spot-modulation will depend on the intrinsic amplitude of the signal in comparison to other sources of stellar or instrumental noise. The amplitude itself is a function of the number, size, contrast, and relative orientations of spots on the surface of stars, which presumably vary over the timescale of spot evolution and the activity cycle of a given star. 

When discussing activity levels, it is not without precedent to invoke a Rossby number-amplitude relation. We define the Rossby number as the rotation period divided by the convective overturn timescale one pressure scale height above the convective boundary in our stellar models, $\textrm{Ro}\equiv P/\tau_{cz}$. \citet{noyes1984} found that Ro appears to set the chromospheric activity in a star. Subsequent papers confirmed the correlation and noted the presence of a ``saturation'' in the relationship at very high activity level. \citet{messina2001} confirmed that photometric variability increases with decreasing Rossby number, and \citet{hartman2009} presented a relation between Rossby number and photometric variability for the young cluster M37, which takes the form of a power law at high Rossby number, and saturates for very low Rossby numbers. Given these empirical results, we use the Rossby number as a model-based predictor of the average magnitude of the starspot-induced variability.  

Figure \ref{ro_patterns} provides a visualization of the value of the Rossby number as a function of rotation period and spectral type in our population model. Under the assumption of a Rossby number-amplitude relation, lines of iso-Ro in this figure are approximately lines of iso-amplitude. Under our assumptions, stars with smaller Ro values have higher amplitudes and are more easily detected in spot modulation. The upper envelope of rotation periods in the dwarfs in Fig. \ref{ro_patterns} is set by the oldest stars in the TRILEGAL model. Above 5500K, the longest period objects are those stars that have already left the MS. 

\begin{deluxetable*}{lcc}
\tablecolumns{3}
\tablecaption{Fitted values of $\textrm{Ro}_{thresh}$.}
\tablehead{
  \colhead{Pop. model} &
  \colhead{$\textrm{Ro}_{thresh}$} &
  \colhead{$\chi^2$}
}
\startdata
Standard \tablenotemark{a} & $2.08^{+0.03}_{-0.04}$ & 73 \\
Pinsonneault et al. (2012) temperatures & $2.01^{+0.04}_{-0.03}$ & 48 \\
van Saders et al. (2016) braking law, $\textrm{Ro}_{crit} = 2.16$ & $2.03^{+0.04}_{-0.05}$ & 79 \\
van Saders et al. (2016) braking law, $\textrm{Ro}_{crit} = 2.16$, 100K $T_{eff}$ errors & $2.00^{+0.04}_{-0.03}$ & 86 \\
van Saders et al. (2016) braking law, varying $\textrm{Ro}_{crit}$, no SGB, 100K $T_{eff}$ errors & $1.98^{+0.02}_{-0.05}$ & 66 \\ 
\enddata
\tablenotetext{a}{The standard fit uses the vanilla population model (with \textit{Kepler} and MMA14 selections) with a simple cut in Rossby number to define the edge, and adds 100K effective temperature errors to the model. Both subgiants and dwarf stars are used in the computation of the distribution edges. We use the effective temperatures from \citet{mathur2017} for the comparison to observations.}
\label{tbl:rothresh_fit}
\end{deluxetable*}

Under this simplified picture, if there exists a threshold in spot modulation amplitude below which period-detection methods become ineffective, this would also correspond to a threshold Rossby number, $\textrm{Ro}_{thresh}$, above which modulation is undetected. It is clear from Figure \ref{ro_patterns} that a truncation of the distribution at a particular threshold value of Ro would result in edge in the observed period distribution at long periods, and that this edge moves to shorter rotation periods as the value of $\textrm{Ro}_{thresh}$ decreases. Notably, at a given value of $\textrm{Ro}_{thresh}$, subgiant stars survive the cuts at longer periods than their main sequence counterparts at similar effective temperatures. This is due to the comparatively deeper convective envelopes in subgiants, meaning that they rotate more slowly at a given Rossby number. At a Rossby threshold of $\textrm{Ro}_{thresh}=2$, for example, the morphology of the upper edge in the period distribution for stars cooler than $\sim5300$K is set entirely by the oldest stars in the population; above $5300$K it is set by the existence of a Rossby threshold, both among MS and SGB stars.

\begin{figure}
        \centerline{\includegraphics[angle=0,scale = 0.9]{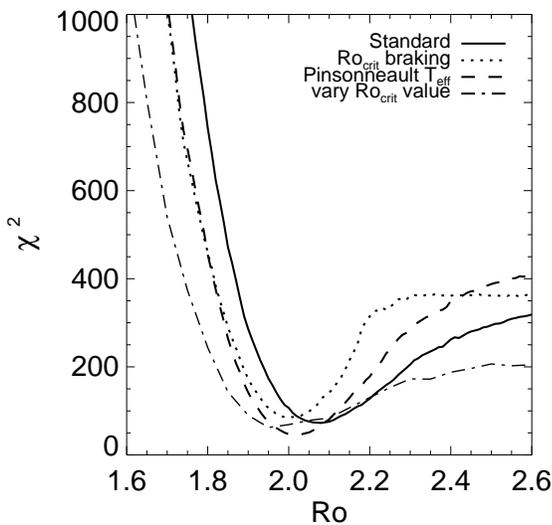}}

        \caption{The Rossby number threshold that best fits the observed long-period edge in the MMA14 rotation distribution. The standard model includes the vanilla braking law of \citet{vansaders2013}, corrections for the MMA14 sample selection and \textit{Kepler} selection function, and uses the \citet{mathur2017} effective temperatures for the comparison to the observed period distribution. The $\chi^2$ curve as a function of Rossby number of detection edge is for the standard model (with 100K temperature errors added) is shown as a solid curve. The same is also reproduced for a comparison using the \citet{pinsonneault2012} temperature scale (with 100K errors, dashed curve) and \citet{vansaders2016} braking law ( with \citet{mathur2017} temperatures and 100K errors, dotted curve). Finally, the dash-dotted curve represents the varying of $\textrm{Ro}_{crit}$ alone, with 100K temperature errors.}
        \label{fig:rodet_fit}
 
\end{figure}

\begin{figure}
        \centerline{\includegraphics[scale = 0.45]{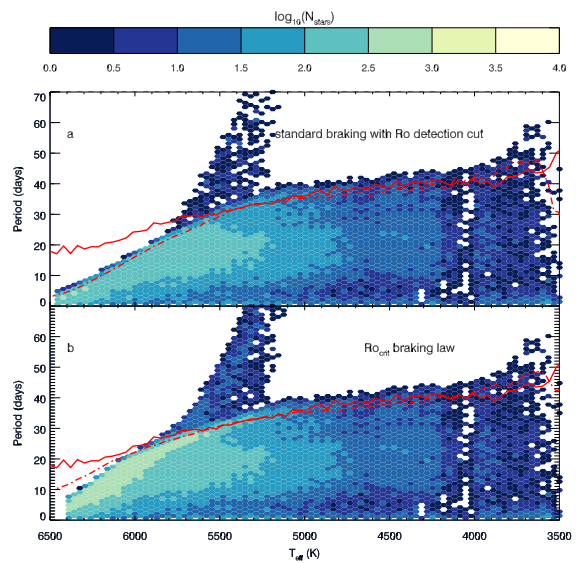}}
        \caption{The effects of detection bias and non-standard braking laws. Panel a: The TRILEGAL population model under a vanilla braking law, corrected for MMA14 cuts and the \textit{Kepler} selection function, where only stars with $\textrm{Ro}< \rothresh$ are detected. Panel b: A population again corrected for MMA14 cuts and \textit{Kepler} selection bias, but now evolved under \citet{vansaders2016} braking law with $\textrm{Ro}_{crit} = \ropick$ In both panels,the solid red line represents the $95^{th}$ percentile period in the observed MMA14 distribution, and the dot-dashed red curve the $95^{th}$ percentile period in the respective model. Populations are shown without injected 100K uncertainties in the effective temperatures.}  
        \label{page_panel_bias}
\end{figure}

We can estimate the value of $\textrm{Ro}_{thresh}$ that best reproduces the observed upper edge in the MMA14 dwarf period distribution. Stars are divided into bins in effective temperature, and we calculate the $95^{th}$ percentile period in each bin. Stars are drawn with replacement in 1000 bootstrap resamplings to estimate the uncertainty on $95^{th}$ percentile period, and the process repeated for a range of $\textrm{Ro}_{thresh}$. We repeat the exercise for the MMA14 sample, and then compared the observed period edge to that predicted in models with a Rossby detection threshold, utilizing the statistic

\begin{equation}
\displaystyle \chi^2 =\sum_{i=1}^{n_{bins}}\frac{\left(P_{edge,mod,i}-P_{edge,obs,i}\right)^2}{\sigma_{P_{edge, mod,i}}^2 + \sigma_{P_{edge, obs,i}}^2}, 
\end{equation}

where $n_{bins}$ represents the number of temperature bins, $P_{edge,mod,i}$ and $P_{edge,obs,i}$ the periods at which 95\% of stars in the sample are more rapidly rotating for the TRILEGAL model and observed MMA14 sample, respectively. The values $\sigma_{edge,obs}$ ad $\sigma_{edge,mod}$ are derived via bootstrap resampling in each bin. Stars are drawn at random with replacement in each bin, and the 95th percentile period value recalculated. Figure \ref{fig:rodet_fit} displays the value of $\chi^2$ as a function of $\textrm{Ro}_{thresh}$. In the corresponding Table \ref{tbl:rothresh_fit}, we provide the best-fit values of the threshold Rossby number with different assumptions about the population model, comparison population, and effective temperature scales. The top panel(a) of Figure \ref{page_panel_bias} demonstrates the impact this Rossby threshold has on our modeled TRILEGAL population detections for $\textrm{Ro}_{thresh} = \rothresh$.  

When searching for an optimal $\textrm{Ro}_{thresh}$, we fit over the effective temperature range of 4500-6000 K, where blending of sources is expected to be less important in determining the observed period distribution (see discussion in Section \ref{blends_and_binaries}). Figure \ref{fig:rodet_fit} displays the fits for various manipulations of the TRILEGAL model. Using the \textit{Kepler} selection corrected samples and adding a $1\sigma$ scatter in effective temperature of 100K \citep{brown2011} to the model values to mimic the KIC temperature uncertainties, we find a best fit Ro threshold of $\textrm{Ro}_{thresh} = \rothresh$. If we instead utilize the \citet{pinsonneault2012} temperatures with 100K temperature errors for the comparison against the TRILEGAL model, the best-fit threshold Ro is $\textrm{Ro}_{thresh} = \pinrothresh$. We explored the effect of removing subgiant stars from the sample before fitting for $\textrm{Ro}_{thresh}$ and find a negligible impact on the inferred value.

\subsection{A Change in the Braking Law}

A modified braking law with a critical Rossby number above which spindown ceases \citep{vansaders2016} can also produce a sharp upper edge in a period distribution. Because the spindown stalls, stars in such a model ``pile up'' at long periods, evolving only very slowly in rotation due to physical changes in their radii and moments of inertia.  In \citet{vansaders2016} this critical threshold was $\textrm{Ro}_{crit} = 2.16$, determined using a small sample of stars with precisely determined asteroseismic ages and spot modulation rotation periods from \textit{Kepler}. The bottom panel of Figure \ref{page_panel_bias} shows our predictions for the period distribution of a TRILEGAL model population evolved under such a critical braking model, with $\textrm{Ro}_{crit}= 2.00$. 

\begin{figure*}
        \centerline{\includegraphics[scale = 0.60]{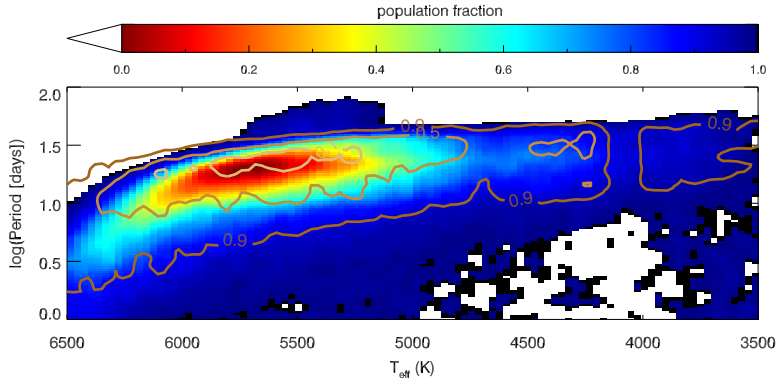}}
        \caption{The comparison between the predicted rotation distribution and observed rotation distribution after accounting for: 1.) MMA14 log(g)-$T_eff$ cuts, 2.) \textit{Kepler} mission bias towards observing FGK dwarfs, 3.) $\textrm{Ro}_{thresh} = \rothresh$ detectability cut, and 4.) 100K uncertainties in the $T_{eff}$ scale. The color-coding of the background can be directly compared with the contours, which enclose 10\%, 50\%, and 90\% of the MMA14 sample. }
        \label{fig:final_fit} 
\end{figure*}

We imagine two scenarios. In the first, we assume that the paucity of long-period stars is entirely due to weakened magnetic braking, and ask which value of $\textrm{Ro}_{crit}$ best reproduces the observed edge by varying the value of $\textrm{Ro}_{crit}$ in the population model; we find a value $\textrm{Ro}_{crit} = \rothreshisrocrit$. In the second scenario, we adopt $\textrm{Ro}_{crit} = 2.16$ from \citet{vansaders2016} but also allow for a detection edge, yielding a best-fit threshold $\textrm{Ro}_{thresh} = \rocritrothresh$. Details regarding the fit values for different temperature scales and underlying assumptions can be found in Table \ref{tbl:rothresh_fit}.

The distribution of stars in the TRILEGAL model, with MMA14 cuts, \textit{Kepler} FGK bias, and effective temperature errors with this threshold ($\textrm{Ro} = \rothresh$) is shown in Figure \ref{fig:final_fit}, in comparison to the MMA14 contours. A fit utilizing the \citet{pinsonneault2012} effective temperature scale with $\textrm{Ro}_{thresh} = \pinrothresh$ provides a qualitatively similar diagram, although subtle differences between the predicted distributions are present due to a shift in effective temperatures between the two scales, particularly in the hot stars.

The TRILEGAL model over-predicts the number of period detections by a factor of $\sim \detfrac$, even with the assumption that there is some Rossby threshold above which stars are not detected in spot modulation. We have not modeled the activity cycles of stars or accounted for the fact that a given star may occasionally present unfavorable spot patterns for detection, and expect that these effects account for at least part of the discrepancy between the number of periods detected in our model versus the observations. We discuss detection fractions further in Section \ref{unmodeled_bias}.

\section{Discussion} \label{sec:conclusions}

\subsection{Rotation-based ages}

Large samples of rotation rates for dwarf stars are of particular interest to the community because of the potential to derive rotation-based ages using gyrochronology. Our population modeling highlights some of the difficulties in the practical application of gyrochronology relations to a stellar population. Dwarf stars must be separated from a mixture of hot, evolved, and blended sources, but differentiating among these populations can be difficult, particularly when surface gravity information is limited. Furthermore, the existence of a Rossby detection threshold (regardless of its origin) systematically biases the rotation-based ages of the stellar ensemble. 

\begin{figure*}
        \centerline{\includegraphics[angle=0,scale = 1.0]{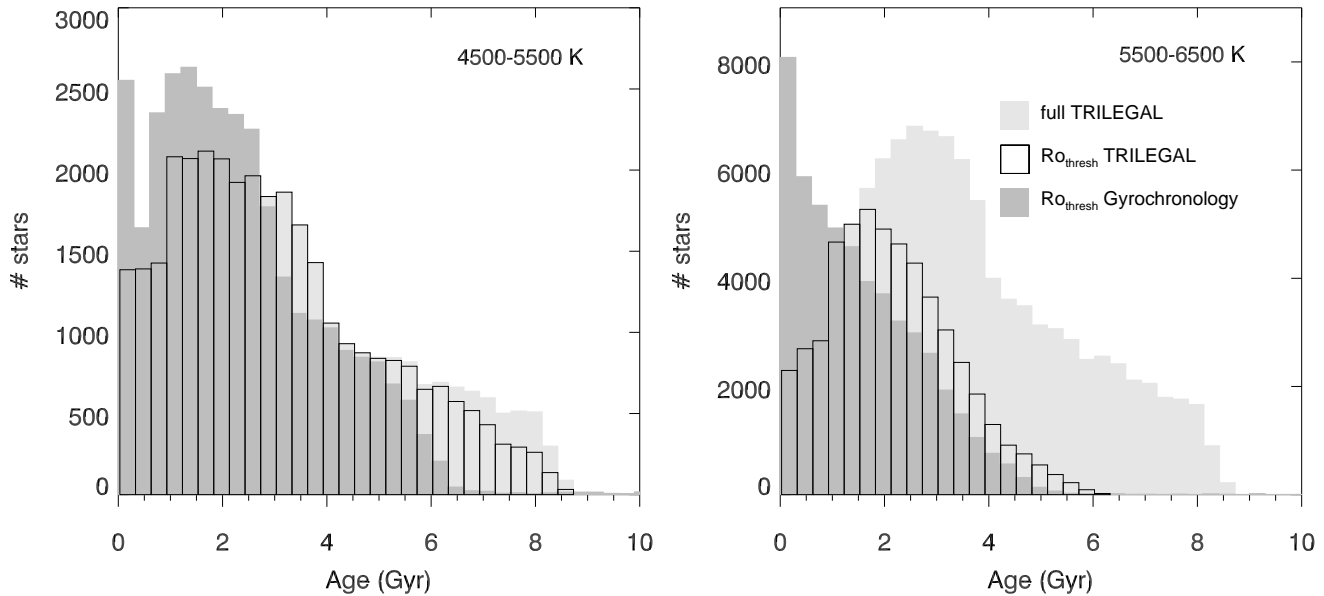}}
        \caption{The comparison between the underlying age distribution in the TRILEGAL model, and age distribution after the stellar sample is subjected to a Rossby detection threshold. Right panel: stars with $5500 < T_{eff}< 6500$K. The light gray histogram shows the TRILEGAL model, with a \textit{Kepler} selection function and $K_p<16$. The open histogram shows the same population and actual model ages, but is limited to those stars with $\textrm{Ro}<\textrm{Ro}_{thresh}$. The dark gray histogram shows the ages one would have inferred using gyrochronology \citep[using the calibration of][]{angus2015}, given the stellar rotation periods and color temperature relationship from \citet{sekiguchi2000}. Left panel: same as the right, but for $4500<T_{eff}<5500$K. For all histograms, only stars with periods $P<70$ days are shown, mimicking the period search in MMA14.}
        \label{age_histo_panels}
 
\end{figure*}

We show in Figure \ref{age_histo_panels} the apparent age distributions that would be inferred from a detection-biased sample. 
In comparison to the full TRILEGAL population (subject only to target selection effects), a population in which only stars with $\textrm{Ro}<\textrm{Ro}_{thresh}$ are detected appears younger; old stars are preferentially missed. This effect is especially pronounced in solar temperature stars and nearly absent in the cooler stars, due to the fact that cool stars reach the threshold Rossby number at ages that often exceed the age of the galactic disk.

\begin{figure}
        \centerline{\includegraphics[angle=0,scale = 1.0]{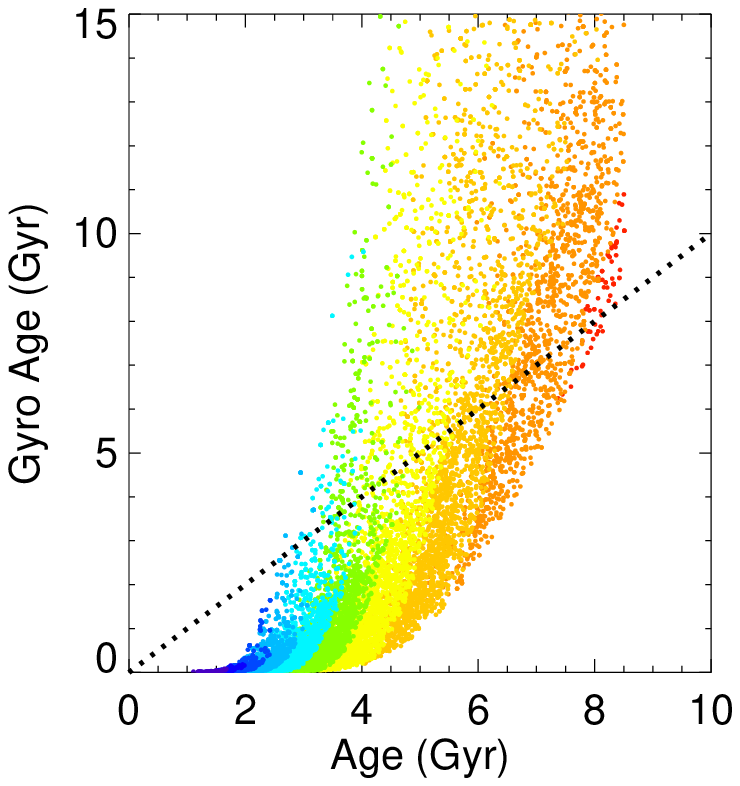}}
        \caption{Standard gyrochronology ages for subgiants, as a function of true age. Subgiants are drawn from the full TRILEGAL sample corrected for \textit{Kepler} bias. Points are color-coded by mass in bins of $0.1 \textrm{M}_{\odot}$ : red points represent stars with $\textrm{M}=0.9-1.0 \textrm{M}_{\odot}$, and the dark blue points stars with $\textrm{M}=1.9-2.0 \textrm{M}_{\odot}$. Massive subgiants have apparent gyrochronological ages younger than their true ages, while low-mass subgiants appear older than their true ages. The dashed black line represents agreement between the true and gyrochronological ages.}
        \label{sgb_ages}
 
\end{figure}

Subgiants also prove to be a confounding population, again particularly in samples of stars near solar temperature. Massive (and thus rapidly rotating) subgiants can be mistaken for young dwarfs, and extremely slowly rotating low-mass subgiants as exceptionally old stars. Figure \ref{sgb_ages} demonstrates the bias: we show the comparison between the true model age of a subgiant, and its apparent age if its period and color were used to infer its age via a standard gyrochronology relationship \citep{angus2015}. This is an area in which Gaia parallaxes will be exceptionally useful: one will be able to reliably discriminate between dwarfs and subgiants, thus avoiding this form of period confusion. 

In both panels of Figure \ref{age_histo_panels} we plot the age distribution one would have inferred were they to take our modeled stars and their periods as truth, and applied the \citet{angus2015} gyrochronology relation. In this scenario, the age distribution is skewed young, for several reasons. First, presuming that stars with $\textrm{Ro}<\textrm{Ro}_{thresh}$ are not detected, old stars are missed, and empirical period-age relations will be subject to the same age bias discussed above. Second, rapidly rotating subgiants (in the absence of reliable subgiant/dwarf discrimination) get mistaken for young stars, causing intermediate-age stars to be tagged as young. Finally, our assumed braking model and the \citet{angus2015} calibration predict slightly different periods for a star of a given age and color, in the sense that our model periods are somewhat shorter. By applying the \citet{angus2015} calibration to our modeled periods, we therefore inject an age bias. This offset is more severe (20\% in period) in the cooler stars, and less pronounced near solar temperatures. The calibration sample in \citet{angus2015} contains essentially no old-star anchors at temperatures cooler than the Sun; it is therefore unsurprising (and not yet particularly concerning) that discrepancies appear between the two models in the coolest stars. Taken together, these features of the rotation distribution require that care is taken in the interpretation of rotation periods: period and color alone are not sufficient to fully understand rotating populations.

\subsection{Origin of the edge: Detection threshold or physical phenomenon?} \label{sec:edge_source}

We have shown that for solar temperature stars, there appear to be far fewer slow rotators in the observed MMA14 sample than we would have predicted with a vanilla stellar population and braking law. We have also shown that we can reproduce this morphology either by assigning a threshold Rossby number above which detections are difficult, by changing the underlying braking law, or by invoking a combination of the two effects. Here we evaluate the various scenarios we have proposed, and suggest observational tests to distinguish among them. 

\subsubsection{Simple detection bias}
Under the assumption that spot modulation amplitude scales with Rossby number \citep[e.g.][]{messina2001,hartman2009}, the existence of an apparent Rossby edge in the population could be the result of a simple detection bias against low amplitude variability. In this scenario, the value of $\textrm{Ro}_{thresh}$ is set largely by the photometric precision of the survey and the sensitivities of the methodologies used to detect spot modulation. Improvements in observations or processing should result in a different inferred value of $\textrm{Ro}_{thresh}$ as stars become detectable at lower amplitudes.  

Some features of the amplitude distribution of the detected rotational modulation are at odds with a simple detection edge. Those stars straddling the apparent edge in the MMA14 period distribution are not the lowest amplitude signals detected, as would be the expectation in a pure detection bias scenario. The median amplitude for a star in the 95th-99th percentile period range in MMA14 in a 100K temperature bin centered on 5800K is $\sim \edgeampsol$ ppm, with a median period of $\sim \edgeperiod$ days.  In contrast, cooler stars are observed with comparable amplitudes out to periods of 40 days, demonstrating that it is possible to detect signals of similar amplitude at longer rotation periods. Hotter stars are observed with amplitudes in the 100s of ppm (albeit with shorter rotation periods), demonstrating that substantially lower amplitude signals are also detectable. 

Furthermore, if we examine the amplitudes of stars at all periods within a 100K bin centered at 5800K, we find that the lowest amplitude signals are not always observed in the longest period stars. In comparison to those stars at the observed period edge ($95-99^{\textrm{th}}$ percentile periods),  $\edgefraclowamp \%$ of stars throughout the bin have lower observed amplitudes. Of those $\edgefraclowamp \%$ of stars, $\edgefraclowampfast \%$ have periods less than 15 days.  The likely presence of blended and contaminated sources provides a potential explanation for low amplitude modulation signals dispersed widely in temperature and period. \textit{Kepler} pixels are large (4 arcsec), and the presence of binary companions or unassociated background stars in the pixel mask used to measure the stellar lightcurve can dilute modulation signals. Sources with diluted amplitudes will not be localized at the apparent detection edge, but rather spread throughout the temperature-period diagram.  

\subsubsection{A change in stellar spottedness}

The origin of the edge could instead be rooted in physical changes in the spottedness and activity of stars as they age. A decrease in the spot coverages, lifetimes, spot sizes, or activity cycles of stars at a given Rossby number could result in an apparent sharp edge in the period distribution. If, for example, the evolution to high Rossby numbers triggered Maunder-like minima in old stars, they would become essentially undetectable in spot modulation. This is particularly interesting in light the \citet{metcalfe2016} suggestion that the Sun is in the midst of a magnetic transition---as evidenced by its unusual magnetic cycle period, the presence of an activity cycle when other stars of the same spectral type and rotation periods show only flat activity, and the shutdown of magnetic braking around solar age suggested by \citet{vansaders2016}. \citet{metcalfe2017} postulates that magnetic cycle periods lengthen as stars pass the critical Rossby number. Although the link between these other proxies of magnetism and the spottedness of stars is not precisely established, the presence of an upper edge to the period distribution is perhaps another symptom of the same underlying shifts in the magnetic character of stars at $\textrm{Ro} \sim 2$.

If this scenario---that the edge is a result of a shift in the physical spot properties of stars---is the sole driver of the morphology of the period edge, future efforts to measure spot modulation should yield an edge in essentially the same location as that observed in MMA14. Other methods of period determination, however, should detect stars at periods beyond the edge. Precise vsini's are a relatively clean way of testing the location of the edge, although they are technically challenging. Ca II H\&K monitoring provides another window into the rotation rates, although there is no guarantee that a shift in spot properties is not accompanied by shifts in the other activity indicators. Asteroseismically inferred rotation periods \citep{davies2015, nielsen2015} provide another relatively clean, if observationally intensive test. In each of these cases, the ability to perform the test is generally limited by the current sample size.  

\subsubsection{A change in stellar-spin down}
\begin{figure}
        \centerline{\includegraphics[angle=0,scale = 1.0]{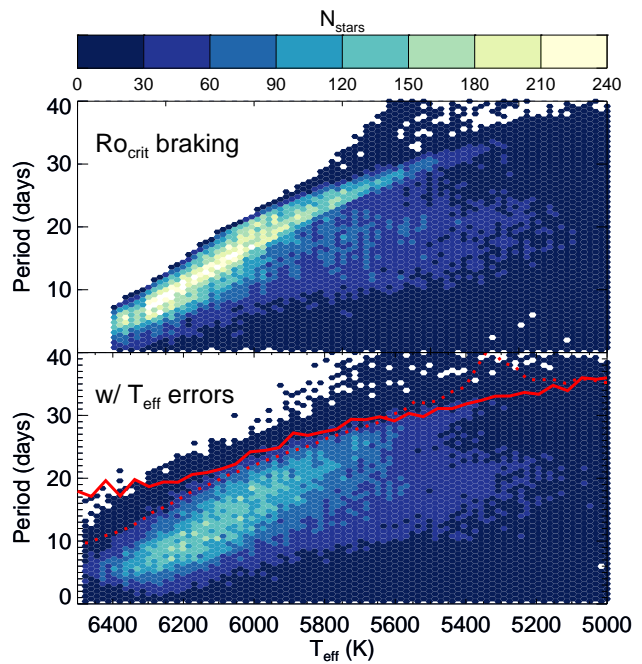}}
        \caption{Top panel: The period distribution assuming a modified braking law of the form in \citet{vansaders2016}, corrected for FGK star selection bias. The overdensity of stars near the long-period edge is a consequence of the shut-down of magnetic braking. Bottom panel: The same model, now with temperatures scrambled by the addition of $\sigma = 100$K $T_{eff}$ errors. Compare to the observed distribution in Figure \ref{MMA14_diagram}.}
        \label{pileup_plot}
 
\end{figure}

In our simulations with the \citet{vansaders2016} braking law, we can produce an edge in the period distribution by shutting down magnetic braking for stars above a critical Rossby number. This explanation has the benefit that the edge in period detections and anomalous spin-down behavior noted in \citet{vansaders2016} and \citet{angus2015} have a common source. However, the hallmark of such a model is a pileup of stars near the edge of the period distribution, which is not seen in the MMA14 sample. However, we show in Figure \ref{pileup_plot} that errors in the observed effective temperatures can effectively smooth this overdensity, erasing the ``smoking gun'' signature of modified braking. Furthermore, if both the braking properties and spottedness of stars undergo a simultaneous shift in their behaviors, it could result in an undetectable pileup. 

If the edge in the period distribution is caused by a change in the underlying braking behavior of stars, it should be present in all diagnostics of rotation period: rotational velocities, Ca H\&K monitoring, and asteroseismology should all arrive at period distributions that agree with the MMA14 period distribution. Improved stellar parameters should sharpen the period distribution and make a pile-up in period evident near the long-period edge, provided that the change in braking is not also accompanied by a change in spot properties that makes long-period stars more challenging to detect. Although we predict a pile-up at long periods, stars do continue to evolve in period as they physically expand in the latter half of the MS while conserving angular momentum: stars can therefore be present above the edge created by a $\textrm{Ro}_{crit}$ braking law, albeit in smaller quantities.

\subsubsection{Unmodeled bias} \label{unmodeled_bias}

\begin{figure}
        \centerline{\includegraphics[angle=0,scale = 1.0]{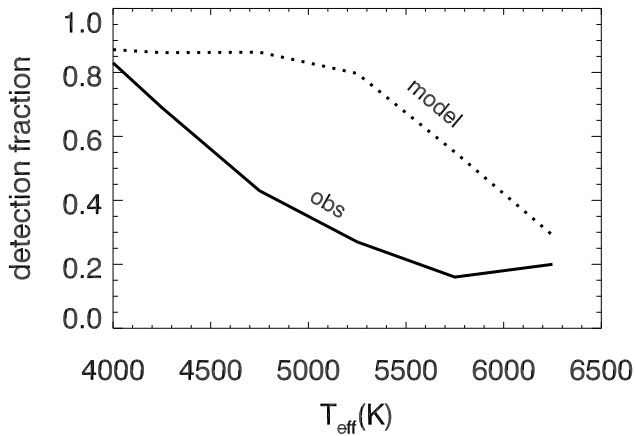}}
        \caption{Observed detection fractions as a function of  effective temperature in comparison to model predictions. The model detection fractions account for the \textit{Kepler} selection bias and assume that stars with $\textrm{Ro}>\textrm{Ro}_{thresh}$ and inclinations above $60^\circ$ are undetectable in spot modulation.}
        \label{det_frac_plot}

\end{figure}

Quite apart from the existence and origin of the long-period edge, there is still detection bias present in MMA14 that we have failed to model. Although our attempts to account for detection bias with a vanilla braking law produce a trend in the fraction of stars detected in spot modulation as a function of spectral type (with the detection fraction declining for hotter stars), it is not of the magnitude seen in MMA14. For example, assuming that all stars have inclinations $i = 0^\circ$, we would expect to detect all stars cooler than $\sim5000$ K and $\soldet \%$ of those with temperatures $5500-6000$K in our standard model with a Rossby detection edge. If we instead assume that all stars host active regions at $30^\circ$ and that when inclined beyond $60^\circ$ spot modulation becomes undetectable, we would expect to detect $\sim \incdetfrac \%$ of those stars cooler than 5000 K, and $\sim \incdetfracedge \%$ at $5500-6000$K. In comparison, MMA14 detected  $\sim60\%$ cooler than 5000K and  $16\%$ of those with temperatures $5500-6000$K. The presence of magnetic cycles and evolving spot patterns is likely responsible for some portion of this discrepancy. Figure \ref{det_frac_plot} plots the detection fraction curves as a function of temperature for both the observed MMA14 sample and our model (assuming both orientation and $\textrm{Ro}_{thresh}$ effects). The observed detection fractions are lower than the model predictions in all temperatures bins, suggesting that we have not adequately modeled all factors that affect detectability. 

\subsubsection{Blends and Binaries} \label{blends_and_binaries}
We comment briefly on the impact of blending and binarity of stars in our model population. Contamination can arise in several flavors: 1) close, synchronized binaries that affect the rotation periods of both stars, 2) wide binaries, in which two separate spot modulation signals are blended within a given lightcurve but are otherwise consistent with single-star evolution, and 3) unassociated blends/contamination, in which light from an unassociated star either dilutes or contributes to the modulation signal. Synchronized binaries will manifest themselves as rapid rotators with periods of a few days, and while they are an important contaminant when searching for young stars, they do not impact our conclusions about the long-period edge. Likewise, the diluting effect of additional light from a companion star may affect the detectability of a given modulation signal and final sample amplitude distributions, but dilution should occur across the entire period-temperature diagram and does not alter the periods themselves. 

The effects of an additional modulation signal from wide binaries and unaffiliated blends can be more subtle. K2 observations of Praesepe and the Pleiades provide a clean laboratory to test the effects of binarity on lightcurves: stars are known to be of the same age, and their rotation periods are generally short and spot modulation of high amplitude. Some $\sim20-25\%$ of stars in the \citet{rebull2016,rebull2017} have multiple highly significant periodicities evident in their lightcurves. In the rapidly rotating M dwarfs, in particular, these multi-period stars are often also photometric binaries \citep{stauffer2016, rebull2016, rebull2017}, suggesting that the two periods correspond to the rotation periods of both components. Furthermore, slowly rotating cluster members are evident above the slow cluster sequence \citep{stauffer2016}; these stars are most likely binaries, where the abnormally slow period is actually that of a more active, more slowly rotating cool companion. A detailed analysis of the confounding effects of binaries and blends on spot modulation signals is a topic of future work and beyond the scope of this article; for now, we provide a word of caution that the interpretation of rotation periods can be complicated by the presence of unrecognized blended or binary companions.

\subsection{The diagnostic power of the period edge in cool stars}

We have discussed the upper edge in the period distribution stars hotter than $\sim 5000$ K extensively; however, in each of these scenarios, the edge in the cooler stars is purely the result of a stellar population with a finite age. Given a particular braking prescription, a population with a finite age will naturally show an upper edge in the period-temperature diagram. Our models reproduce the edge in the cool stars well, regardless of any complicating $\textrm{Ro}_{thresh}$ or $\textrm{Ro}_{crit}$ modifications. Because this upper edge is set only by the presence of old stars and the strength of a vanilla magnetic braking prescription, it can provide interesting physical constraints. In a population that is not well understood, standard braking laws can provide an estimate of the oldest stars in the population. If instead we have a population whose age distribution we think is well-understood (in principle easier than providing precise ages for individual stars), this upper edge is yet another constraint on the strength of magnetic braking, and can be used as an old-star calibration point in period-age relations.

\subsection{Unexplained features in the rotation distribution}
There are a number of features in the MMA14 distribution that we have thus far been unable to explain with our population model: namely the ``dip'' in the M star rotation periods, the bimodal period distribution observed in the cool stars, and sharp lower edge to the period distribution (see Figure \ref{MMA14_diagram}). \citet{mcquillan2013} and \citet{mcquillan2014} suggested that the observed period bimodality may be due variations in the age distribution of stars in \textit{Kepler} field. In particular, they invoke bursty star formation in the solar neighborhood to explain the bimodality in cool stars, while also explaining the disappearance of the bimodality at temperatures hotter than 4500 K with the observation that hotter stars can be seen to larger distances. \citet{davenport2017} found, using the subset of the MMA14 sample with Gaia parallaxes to isolate dwarf stars, that this bimodality extends through into the hot stars in the MMA14 sample. This result is in agreement with our prediction that the period-temperature diagram is contaminated with crossing subgiant stars for temperatures above $\sim5800$K. The age distribution in the TRILEGAL model, as is, does not include this bursty, localized star formation, and displays no such bimodality. An artificially added burst of star formation with a duration of $\sim 0.5$ Gyr can qualitatively reproduce the short-period feature observed in MMA14, and a sharp lower edge to the period distribution.  If we assume that stars closer than 500 pc have a bursty star formation history, we can also reproduce a bimodality that fades at $\sim4500$ K. Notably, the morphology is not reproduced entirely successfully: the braking law produces a ``downturn'' in the observed period of the bimodality for the coolest stars, due to the presence of stars with rotation rates in the saturated domain for angular momentum loss.

\subsection{The future: Gaia, TESS, and K2}

We make a number of predictions directly testable with upcoming data from space missions. 

\begin{enumerate}
\item{Gaia will enable efficient subgiant vs. dwarf discrimination. \citet{davenport2017} has already demonstrated the apparent period bimodality extended into the hot stars by selecting dwarfs with Gaia DR1; we argue that this is primarily because the signature was obscured by ``contaminating" subgiants in the absence of precision surface gravity/luminosity constraints. Gaia DR2 will provide precision parallaxes for every star in MMA14, providing a direct test of whether populations enumerated in Figure \ref{page_panel_selection} are present.}

\item{K2 and TESS asteroseismology will provide measurements of the near-surface rotation from rotational splitting of the oscillation frequencies that do not depend on spot-modulation. This will provide a test of whether the long-period edge is due to detection bias, modified braking, or a shift in stellar spottedness.}

\item{Spot-modulation rotation periods from K2 and TESS will probe different regions of the sky and thus different stellar populations. If the long-period edge to the period distribution is tied to the physics of magnetic braking or stellar spottedness, it will be universally observed. If, on the other hand, it is a pathological feature of the stellar population in the \textit{Kepler} field, it will not be observed along all other sight-lines.  }
\end{enumerate}

\subsection{Conclusions}
By coupling a TRILEGAL model of the galaxy with theoretical rotation models, we can largely reproduce the observed distribution of rotation periods in MMA14. In order to do this we must account for selection biases, and in particular, institute a Rossby threshold above which we assume that stars are not detected, or weaken the magnetic braking law. The best-fit Rossby detection threshold of $\textrm{Ro}_{thresh}= \rothresh$ naturally matches the shape of the observed drop-off of detections of long-period stars in the MMA14 sample. We argue that this edge is likely physical in origin, rather than a pure detection edge, and is the result of either an abrupt change in the spottedness of stars or their braking behavior. Our rotation modeling emphasizes the fact that we expect to see a mixed population of hot, evolved and cool main sequence stars, and that these populations overlap in period-$T_{eff}$ space, making the interpretation of rotation periods in the context of gyrochronology more challenging. This mixture of populations and Rossby detection threshold conspire to severely bias the rotation-based ages of stars in this sample, a feature which must be appreciated before realistic age distributions can be extracted from this, or any other large rotational dataset.

\acknowledgments  
We thank Travis Metcalfe, Benjamin Shappee, Leo Girardi, and Boaz Katz for stimulating discussions. J.v.S. acknowledges the support of a Carnegie-Princeton Fellowship. MP acknowledges support from NASA grant NNX15AF13G.  This research was supported in part by an Universidad De Atacama grant (DIUDA2015).


\end{document}